# Quantum effects in many-body gravitating systems

V. A. Golovko

Moscow State Evening Metallurgical Institute, Lefortovsky Val 26, Moscow 111250

Russia

Abstract

A hierarchy of equations for equilibrium reduced density matrices obtained earlier is used to consider systems of spinless bosons bound by forces of gravity alone. The systems are assumed to be at absolute zero of temperature under conditions of Bose condensation. In this case, a peculiar interplay of quantum effects and of very weak gravitational interaction between microparticles occurs. As a result, there can form spatially-bounded equilibrium structures macroscopic in size, both immobile and rotating. The size of a structure is inversely related to the number of particles in the structure. When the number of particles is relatively small the size can be enormous, whereas if this number equals Avogadro's number the radius of the structure is about 30 cm in the case that the structure consists of hydrogen atoms. The rotating objects have the form of rings and exhibit superfluidity. An atmosphere that can be captured by tiny celestial bodies from the ambient medium is considered too. The thickness of the atmosphere decreases as its mass increases. If short-range intermolecular forces are taken into account, the results obtained hold for excited states whose lifetime can however be very long. The results of the paper can be utilized for explaining the first stage of formation of celestial bodies from interstellar and even intergalactic gases.

**1. Introduction**

Gravitational interaction between atomic particles is extremely week and is usually disregarded. At the same time, if forces of gravity alone acted between the particles, one would obtain rather curious results. In classical statistical mechanics, at absolute zero of temperature ($T = 0$) the particles would coalesce into a point, if attractive forces alone were involved, regardless of the strength of the attraction. In quantum mechanics, the particles cannot coalesce because of the uncertainty principle. If one takes, for example, the Bohr radius $a = \hbar^2/(me^2)$ and replaces $e^2$ by $Gm^2$ ($G$ is the gravitation constant and $m$ is the particle mass) one will obtain $a_G = \hbar^2/(Gm^3) \sim 10^{19}$ km, once the mass of a hydrogen atom is put for $m$. We see a drastic difference between the classical and quantum cases: $a_G = 0$ if $\hbar = 0$, whereas $a_G \sim 10^{19}$ km in the quantum case. The latter value of $a_G$ is, of course, relevant to a system consisting of only two atoms. The aim in this paper is to investigate the size and structure of quantum systems containing a great number $N$ of particles that interact via the forces of gravity (the $T = 0$ case is implied). We shall see that the size of the systems that can be immobile or rotating depends essentially upon $N$. For example, a mole of atomic hydrogen held solely by gravitational forces of its own should occupy a sphere of about 30 cm in radius instead of $10^{19}$ km. It is to be added that in numerous statistical studies on gravitating systems (for a review see [1]) only the classical case is considered for the most part.

The question arises as to the conditions under which the gravitational interaction between atoms or molecules can play a leading role (for brevity, we shall speak of atoms). Familiar short-range interatomic forces can be neglected as compared to the long-range gravitational forces, if distances between the atoms are sufficiently great, that is to say, if the density of a gas is sufficiently low. Among the former forces, the longest range is characteristic of the van der Waals forces whose potential in the case of two hydrogen atoms is of the form [2,3]

$$K_W(r) = -\frac{6.50\, e^2 a_B^5}{r^6}, \qquad (1.1)$$

where $a_B = \hbar^2/(m_e e^2)$ is the ordinary Bohr radius and $m_e$ is the electron mass. If one computes the force corresponding to (1.1) and the one corresponding to the Newtonian potential

$$K(r) = -\frac{Gm^2}{r}, \qquad (1.2)$$

one will see that the forces become equal to each other at $r = r_0$ and the relevant density $\rho_0$ is

$$\rho_0 = \frac{1}{r_0^3} = \frac{1}{a_B^3}\left(\frac{Gm^2}{39.0\, e^2}\right)^{3/5}. \qquad (1.3)$$

This gives $r_0 = 0.18$ cm and $\rho_0 = 166$ atoms/cm$^3$ for hydrogen. By virtue of the rapid decrease of $K_W(r)$ as $r$ increases, it is safe to say that when the average density of a hydrogen gas is less than $10^2$ atoms/cm$^3$ there remains only the gravitational interaction between the atoms. In the case of other atoms or molecules, approximate formulae and experimental data presented in [3] enable one to estimate the relevant value of $r_0$ and $\rho_0$ in like fashion. It must be stressed that, even if $\rho > \rho_0$, the influence of the gravitational interaction may still be essential because, in the case of short-range forces, only a limited number of particles act on a given particle; while in the case of long-range forces a great number of particles act on the same particle.

In our Galaxy, the density of interstellar neutral hydrogen is about 1 atom/cm$^3$ dropping to several hundredths of atom/cm$^3$ at the periphery. The intergalactic gas and the substance of some extragalactic gaseous nebulae have densities that are significantly lower. Hence, in all these cases the atoms interact with one another essentially via gravity. It should be emphasized that we imply those regions of outer space where the temperature is close to zero because the results of the present study are applicable directly only to a gas at $T = 0$.

Seeing that the interstellar gas consists, for the most part, of atomic hydrogen whereas hydrogen atoms in the ground state are spinless bosons, in this paper we make use of the approach developed in [4] just for quantum systems of spinless bosons at thermodynamic equilibrium. At low temperatures, a phenomenon analogous to the Bose condensation in an ideal gas should happen. Modification of the results of [4] with account taken of the phenomenon was performed in [5] whose equations form the basis of the present study. It is worthy of remark that in [6,7] the methods of [4,5] are extended to the case of non-zero spin particles, including fermions. As long as the gravitational fields are assumed to be weak in this paper, resorting to the general theory of relativity is not required; so we shall restrict ourselves to Newton's theory of gravity.

## 2. Basic equations

The approach developed in [4,5] makes use of $s$-particle reduced density matrices $R_s(\mathbf{x}_s, \mathbf{x}'_s)$ with $s = 1, 2, \ldots, N$ where $\mathbf{x}_s$ denotes a set of coordinates $\mathbf{r}_1, \mathbf{r}_2, \ldots, \mathbf{r}_s$ and $N$ is the total number of particles in the system. At thermodynamic equilibrium, a hierarchy of equations can be obtained, which contains only diagonal elements of the density matrices

$$\rho_s(\mathbf{x}_s) = R_s(\mathbf{x}_s, \mathbf{x}_s). \qquad (2.1)$$

If Bose particles are dealt with, at low temperatures a number of particles prove to be in a special state and form a condensate. As long as in this paper the $T = 0$ case is considered, we shall presume that the condensate comprises all particles so that the density matrices $R_s(\mathbf{x}_s, \mathbf{x}'_s)$ coincide with the condensate parts $R_s^{(c)}(\mathbf{x}_s, \mathbf{x}'_s)$, while the normal parts $R_s^{(n)}(\mathbf{x}_s, \mathbf{x}'_s)$ are null. Taking $R_s^{(c)}(\mathbf{x}_s, \mathbf{x}'_s)$ of [5] we have, therefore,

$$R_s(\mathbf{x}_s, \mathbf{x}'_s) = R_s^{(c)}(\mathbf{x}_s, \mathbf{x}'_s) = \varphi_s(\mathbf{x}_s) \varphi_s^*(\mathbf{x}'_s), \qquad (2.2)$$

the functions $\varphi_s(\mathbf{x}_s)$ being found from the equations

$$\frac{\hbar^2}{2m} \sum_{j=1}^{s} \nabla_j^2 \varphi_s(\mathbf{x}_s) + [\varepsilon_{(s)} - U_s(\mathbf{x}_s)] \varphi_s(\mathbf{x}_s) = 0, \qquad (2.3)$$

where $m$ is the particle mass. The effective potentials $U_s(\mathbf{x}_s)$ are determined by the equations

$$\rho_s(\mathbf{x}_s) \nabla_1 U_s(\mathbf{x}_s) = \rho_s(\mathbf{x}_s) \nabla_1 \left[ \sum_{j=2}^{s} K(|\mathbf{r}_1 - \mathbf{r}_j|) + V^{(e)}(\mathbf{r}_1) \right] + \int \rho_{s+1}(\mathbf{x}_{s+1}) \nabla_1 K(|\mathbf{r}_1 - \mathbf{r}_{s+1}|) d\mathbf{r}_{s+1}, \qquad (2.4)$$

in which $K(|\mathbf{r}_i - \mathbf{r}_j|)$ is a two-body potential that describes interaction between the particles, and $V^{(e)}(\mathbf{r})$ is an external potential. Inasmuch as we have now, because of (2.1) and (2.2), that

$$\rho_s(\mathbf{x}_s) = |\varphi_s(\mathbf{x}_s)|^2, \qquad (2.5)$$

equations (2.3) and (2.4) represent a hierarchy of equations ($s = 1, 2, \ldots$) in which only the diagonal elements $\rho_s(\mathbf{x}_s)$ of the density matrices figure rather than the full density matrices.

The quantities $\varepsilon_{(s)}$ of (2.3) with a dimension of energy are specified by the requirement that $\rho_s(\mathbf{x}_s)$ must be interrelated by

$$(N - s + 1)\rho_{s-1}(\mathbf{x}_{s-1}) = \int \rho_s(\mathbf{x}_s) d\mathbf{r}_s. \qquad (2.6)$$

In [5] it was shown that, in the case of uniform media, this last interrelation enables one to express the quantities $\varepsilon_{(s)}$ in terms of $\varepsilon_{(1)}$ alone. In what follows, only $\varepsilon_{(1)}$ is needed.

In this paper, we shall investigate the number density $\rho_1(\mathbf{r}_1)$ and we shall omit the subscript 1 of $\rho_1$, $\mathbf{r}_1$, $\varphi_1$, $U_1$ and $\varepsilon_{(1)}$. In this case equation (2.3) with $s = 1$ takes the form

$$\frac{\hbar^2}{2m} \nabla^2 \varphi(\mathbf{r}) + [\varepsilon - U(\mathbf{r})] \varphi(\mathbf{r}) = 0, \qquad (2.7)$$

whereas in the absence of the external field [1] equation (2.4) yields

$$\rho(\mathbf{r}) \nabla U(\mathbf{r}) = \int \rho_2(\mathbf{r}, \mathbf{r}') \nabla K(|\mathbf{r} - \mathbf{r}'|) d\mathbf{r}', \qquad (2.8)$$

where $\rho_2(\mathbf{r}, \mathbf{r}')$ is a pair density matrix. From (2.5) it follows that

$$\rho(\mathbf{r}) = |\varphi(\mathbf{r})|^2, \qquad (2.9)$$

and, besides, the density $\rho(\mathbf{r})$ should satisfy the normalization condition [4]

---

[1] If the object under study is relatively small in size, the gravitational potential due to the Galaxy can be considered to be a constant within the object. The constant can be incorporated in $\varepsilon$, which is implied hereinafter.

$$\int_V \rho(\mathbf{r}) \, d\mathbf{r} = N, \tag{2.10}$$

in which the integration is carried out over the volume $V$ occupied by the system (this is implied for all space integrals throughout the paper).

In studies of dilute systems in which long-range forces act, good use is made of Vlasov's approximation for the pair distribution function [8]. In our case the approximation implies that

$$\rho_2(\mathbf{r}_1,\mathbf{r}_2) = \rho(\mathbf{r}_1)\,\rho(\mathbf{r}_2). \tag{2.11}$$

Once (2.11) is inserted into equation (2.8), the equation is readily integrated to give

$$U(\mathbf{r}) = \int K(|\mathbf{r}-\mathbf{r}'|)\rho(\mathbf{r}') \, d\mathbf{r}', \tag{2.12}$$

where the integration constant is chosen such that $U(\mathbf{r}) \to 0$ as $|\mathbf{r}| \to \infty$.

If solely forces of gravity act in the system, then the potential $K(|\mathbf{r}-\mathbf{r}'|)$ is given by (1.2). In this event equation (2.12) can be recast in a differential form, if use is made of the Laplacian operator $\nabla^2$. With account taken of (2.9) the equation assumes the form

$$\nabla^2 U(\mathbf{r}) = 4\pi G m^2 |\varphi(\mathbf{r})|^2. \tag{2.13}$$

Equations (2.7) and (2.13), together with (2.9) and (2.10), constitute the equations that form the basis for the following investigation. It is worth noting that equation (2.7) bears close resemblance to the Schrödinger equation, if one regards $\varphi(\mathbf{r})$ as a wave function and determines the potential $U(\mathbf{r})$ in a self-consistent way according to (2.13). This resemblance is, however, purely outward, though useful for qualitative analysis. In actual fact, the genuine wave function is $\Psi(\mathbf{r}_1,\ldots,\mathbf{r}_N, t)$. Equation (2.7) results from integrating the relevant density matrix over all variables except $\mathbf{r}_1$; and the equation describes only a state of thermodynamic equilibrium, provided fluctuations that depend on time $t$ can be disregarded [4,5]. In particular, the quantity $\varepsilon$ in (2.7) does not represent the energy of a particle. The energy $E$ of the system is given by equation (3.6) of [4] that can, on account of (2.2), (2.11) and (2.12), be reduced to

$$E = -\frac{\hbar^2}{2m}\int \varphi^*(\mathbf{r})\nabla^2 \varphi(\mathbf{r}) \, d\mathbf{r} + \frac{1}{2}\int U(\mathbf{r})\rho(\mathbf{r}) \, d\mathbf{r}. \tag{2.14}$$

Substituting (2.7) and making use of (2.10) leads to

$$E = N\varepsilon - \tfrac{1}{2}\int U(\mathbf{r})|\varphi(\mathbf{r})|^2 \, d\mathbf{r}. \tag{2.15}$$

In what follows it is convenient to employ dimensionless quantities. We introduce a parameter $l_0$ with a dimension of length and a parameter $\varphi_0$ that defines the dimension of $\varphi$, upon writing $\mathbf{r} = l_0 \tilde{\mathbf{r}}$ and $\varphi = \varphi_0 f$ with dimensionless quantities $\tilde{\mathbf{r}}$ and $f$. Equation (2.7) acquires now the form

$$\tilde{\nabla}^2 f(\tilde{\mathbf{r}}) + [\tilde{\varepsilon} - u(\tilde{\mathbf{r}})] f(\tilde{\mathbf{r}}) = 0, \tag{2.16}$$

where $\tilde{\nabla} = \partial / \partial \tilde{\mathbf{r}}$, and the dimensionless $\tilde{\varepsilon}$ and $u$ are defined by

$$\varepsilon = \frac{\hbar^2}{2ml_0^2} \tilde{\varepsilon}, \quad U = \frac{\hbar^2}{2ml_0^2} u. \tag{2.17}$$

Equation (2.13) transforms to the following dimensionless form

$$\tilde{\nabla}^2 u(\tilde{\mathbf{r}}) = 4\pi |f(\tilde{\mathbf{r}})|^2 \tag{2.18}$$

and gives a relation between $\varphi_0$ and $l_0$:

$$\varphi_0^2 = \frac{\hbar^2}{2Gm^3 l_0^4}. \tag{2.19}$$

If one links $\varphi_0$ and $l_0$ by another relation, namely, $l_0^3 \varphi_0^2 = N$, equation (2.10) becomes

$$\int_{\tilde{V}} |f(\tilde{\mathbf{r}})|^2 d\tilde{\mathbf{r}} = 1. \tag{2.20}$$

On account of the second relation between $\varphi_0$ and $l_0$, from (2.19) one has finally

$$l_0 = \frac{\hbar^2}{2Gm^3 N} = \frac{\hbar^2}{2Gm^2 M}, \tag{2.21}$$

where the mass of the system $M = Nm$ has been introduced. The energy of the system of (2.15) can also be expressed in terms of a dimensionless quantity $\tilde{E}$:

$$E = \frac{2G^2 m^5 N^3}{\hbar^2} \tilde{E}, \quad \tilde{E} = \tilde{\varepsilon} - \tfrac{1}{2} \int u(\tilde{\mathbf{r}}) |f(\tilde{\mathbf{r}})|^2 d\tilde{\mathbf{r}}. \tag{2.22}$$

It is instructive to recast (2.21) in another form upon introducing the Compton wavelength of the system $\Lambda_C = \hbar/Mc$ and the Planck mass $m_P = \sqrt{\hbar c / G} \approx 2.2 \cdot 10^{-5}$ g:

$$l_0 = \frac{1}{2} \left( \frac{m_P}{m} \right)^2 \Lambda_C. \tag{2.23}$$

The Compton wavelength for a macroscopic system is extremely small. Only if the constituents of the system are microscopic particles, does the factor $(m_P/m)^2$ lead to a reasonable value of $l_0$. For example, in the case of hydrogen $m_P/m \sim 10^{19}$. Note also that, if $N$ is small, the length $l_0$ is of the same order of magnitude as $a_G$ discussed in Introduction.

In the next section we shall see that in the case of spherical structures their radius is of the order $\tilde{R} = 10$, which amounts to saying that in ordinary units their radius will be of the order $R = 10 l_0$. As a concrete example, we shall assume that the system is composed of hydrogen atoms (see Introduction). Taking the mass of a hydrogen atom for $m$ in (2.21), we get the following estimate for the radius of the system

$$R \approx \frac{1.78}{N} 10^{23} \quad \text{(in metres)}. \tag{2.24}$$

It is reasonable to believe that statistical laws manifest themselves in systems with a number of particles of order $N = 100$ and more [9]. Setting $N = 100$ in (2.24), one gets a huge size of order $10^5$ light years, which is comparable with the size of the Galaxy. If one takes the radius of our solar system (the mean distance between the Sun and Pluto) as $R$ in (2.24), one will see that a system with such dimensions will be made up of about $3 \cdot 10^{10}$ hydrogen atoms.

It is informative to estimate the average number density $\bar{\rho}$ for the systems in question. Since the volume of the system is $V = 4\pi R^3/3$, equation (2.24) yields

$$\bar{\rho} = \frac{N}{V} = 4.25 \cdot 10^{-71} N^4 \left( \frac{1}{\text{m}^3} \right). \tag{2.25}$$

As mentioned in Introduction, in the Galaxy the density of interstellar neutral hydrogen is about 1 atom/cm$^3$ dropping to several hundredths of atom/cm$^3$ at the periphery. Substituting these values into (2.25) we see that a bound system whose average density is within this range will contain about $10^{19}$ atoms or slightly less, while its radius will be of the order 10 km or somewhat more, according to (2.24). In the above example of the solar system-sized object, the average density is as low as 1 atom per $3 \cdot 10^{19}$ km$^3$. This situation may occur only in a very deserted region of intergalactic space; in which case the system composed of about $3 \cdot 10^{10}$ particles will occupy a vast volume.

Yet another example is worthy of consideration. Let us calculate the radius of a sphere in which one mole of atomic hydrogen (the number of atoms equal to Avogadro's number) can be kept by its own forces of gravity. Equation (2.24) gives $R \approx 30$ cm in this case. This example is also instructive because it shows that, notwithstanding the extreme weakness of the gravitational interaction between the hydrogen atoms, bound objects of small size may form, even if the familiar interatomic forces are discarded. According to (2.21) the size of an object is rather sensitive to the mass of its constituents. For example, a mole of molecular oxygen held by the forces of gravity alone would occupy a sphere of about $10^{-3}$ cm in radius.

## 3. Spherically symmetric structures

We turn now to analysis and solution of equations (2.16) and (2.18). In order to simplify formulae we shall omit the tilde over **r** and $\varepsilon$, keeping in mind that the functions $f$ and $u$ depend in fact upon $\tilde{\mathbf{r}}$ and $\tilde{\varepsilon}$. We start with spherically symmetric solutions when $f$ is real-valued. Upon writing the Laplacian in spherical coordinates, from (2.16) and (2.18) one has

$$\frac{d^2 f}{dr^2} + \frac{2}{r}\frac{df}{dr} + (\varepsilon - u)f = 0, \qquad \frac{d^2 u}{dr^2} + \frac{2}{r}\frac{du}{dr} - 4\pi f^2 = 0. \tag{3.1}$$

As long as no boundary is implied, the integration in (2.20) is to be extended over all space:

$$4\pi \int_0^\infty r^2 f^2(r)\, dr = 1. \tag{3.2}$$

The nonlinear equations of (3.1) do not lend themselves to solving analytically, and therefore recourse to numerical methods is needed. It is worthwhile to first perform a preliminary analysis of the equations. It is convenient to incorporate the constant $\varepsilon$ into $u(r)$ by introducing a new function $w(r) = u(r) - \varepsilon$. Then the equations of (3.1) acquire the form

$$\frac{d^2 f}{dr^2} + \frac{2}{r}\frac{df}{dr} - w f = 0, \qquad \frac{d^2 w}{dr^2} + \frac{2}{r}\frac{dw}{dr} - 4\pi f^2 = 0. \tag{3.3}$$

When carrying out numerical calculation, the matter is complicated by the fact that one has to take account of the condition (3.2), which is not trivial because of the nonlinearity of the equations. One can however circumvent the complication, owing to the following property of the equations of (3.3) that can readily be checked. If the equations have a solution $[f(r), w(r)]$, then they will admit another solution of the form

$$f_c = \frac{1}{C^2} f\!\left(\frac{r}{C}\right), \qquad w_c = \frac{1}{C^2} w\!\left(\frac{r}{C}\right) \tag{3.4}$$

with an arbitrary constant $C$, while the integral of the type that figures in (3.2) will be divided by $C$ if the new function $f_c(r)$ is substituted. For this reason, one may look for a solution of the equations without taking care to satisfy (3.2). Upon finding the solution, one ought to calculate the integral on the left of (3.2) and to carry out the transformation of (3.4), taking the calculated value of the integral for $C$. The new function $f_c$ will obey the condition of (3.2).

Let us find out the number of arbitrary constants that determine a solution to the equations of (3.3) that is regular at $r = 0$. To this end, one should seek the solution in terms of series in powers of $r$. The series will contain only even powers of $r$, and one will see that all terms of the series will be expressed uniquely via $f(0)$ and $u(0)$, which amounts to saying that there are two arbitrary constants, while $df/dr = dw/dr = 0$ at $r = 0$.

For numerical calculation, the equations of (3.3) were rewritten in the form of a set of four differential equations of first order; and the set was solved with the help of the well-known Runge-Kutta method. Owing to the property of (3.4), one of the two arbitrary constants may be chosen at will. In the course of the calculation, we set $f(0) = 1$ (the sign of $f$ plays no role for the equations), and by the trial-and-error method we looked for a value of $w(0)$ such that $f(r) \to 0$ as $r \to \infty$. The solution obtained was transformed, on account of (3.4), so that (3.2) was fulfilled. Reverting to the equations of (3.1), the value of ε can be found with use made of the fact that $w(r) \to -\varepsilon$ as $r \to \infty$, since $u(r) \to 0$ in this limit (more precisely $u(r) \to -1/r$ if (3.2) holds). The energy of the structure $\tilde{E}$ can be computed with the help of (2.22). However, when no spatial boundaries are implied, a simple relation $\tilde{E} = \tilde{\varepsilon}/3$ holds, which can be proven upon reducing (2.22) with use made of (2.16) and (2.18).

The equations of (3.1) have different solutions, and the parameters of the first three of them are listed in table 1. Solution 1 has no zeros, solution 2 has one zero, solution 3 has two zeros (figure 1); there are solutions with three zeros and more. Since the density $\rho(r) = f^2(r)$, solution 1 corresponds to a sphere, solution 2 corresponds to a sphere surrounded by a spherical layer; in the case of other solutions there is a sphere with several spherical layers.

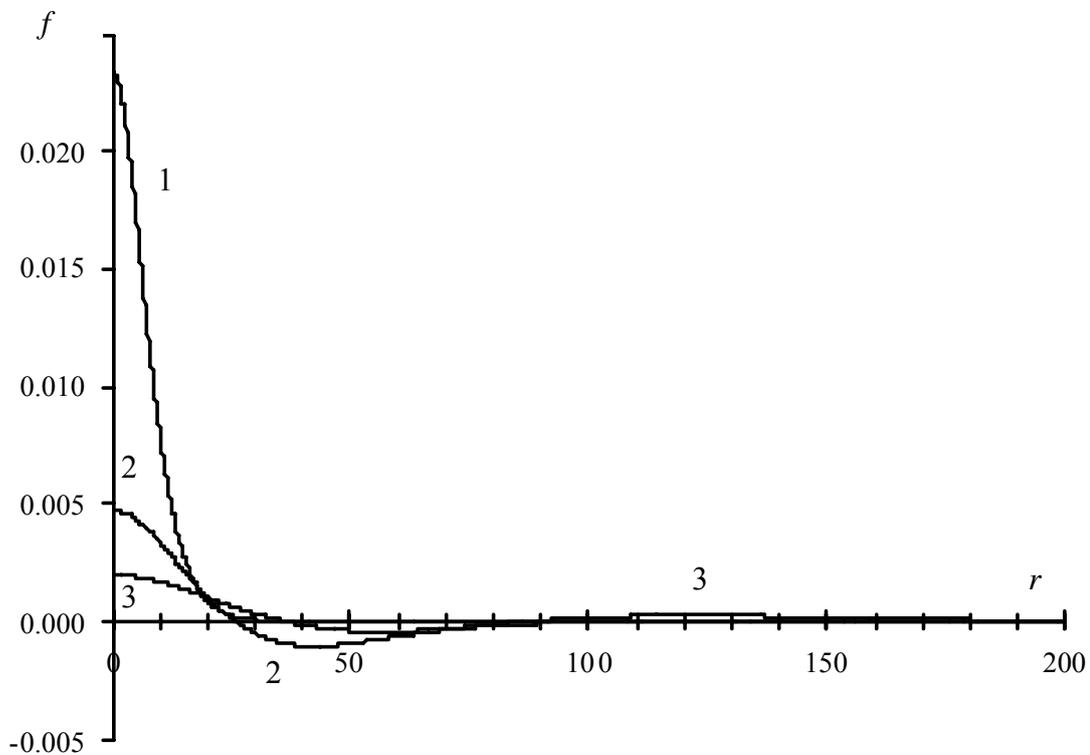

**Figure 1.** Function $f(r)$ for spherically symmetric solutions. The numeration of the solutions corresponds to table 1.

**Table 1**. Parameters of the spherically symmetric solutions.

| Solution | $f(0)$ | $u(0)$ | $\tilde{\varepsilon}$ | $\tilde{E}$ | $\tilde{R}$ |
|---|---|---|---|---|---|
| 1 | $2.345 \cdot 10^{-2}$ | $-0,1577$ | $-8.138 \cdot 10^{-2}$ | $-2.713 \cdot 10^{-2}$ | 10.4 |
| 2 | $4.741 \cdot 10^{-3}$ | $-3.573 \cdot 10^{-2}$ | $-1.540 \cdot 10^{-2}$ | $-5.133 \cdot 10^{-3}$ | 70.8 |
| 3 | $1.980 \cdot 10^{-3}$ | $-1.570 \cdot 10^{-2}$ | $-6.263 \cdot 10^{-3}$ | $-2.088 \cdot 10^{-3}$ | 174.2 |

The equilibrium state of the system under study is that whose energy is minimal. According to table 1, this is the state described by solution 1. Having regard to (2.22) the energy of this last state expressed in ordinary units is

$$E = -\frac{G^2 m^5 N^3}{7.37 \hbar^2}. \tag{3.5}$$

This formula agrees with the estimate of the ground-state energy of a gravitating system given in [1] (the numerical coefficient of [1] is $\approx 1/8$ instead of 1/7.37 in (3.5)).

The radius of the structure may be conventionally defined as a distance from the centre (or from the last maximum) to the point where the density $\rho(r)$ diminishes by a factor of 10. The radii of the structures $\tilde{R}$ are listed in table 1 as well. The radius relevant to solution 1 was utilized for the estimate of (2.24).

## 4. Rotating structures

If particles had a nonzero total angular momentum, then the bound system that arises should also possess an angular momentum, that is, the system should rotate. To begin with, let us deduce a formula for calculation of the angular momentum of a system in the framework of the approach used. If the system is described by a wave function $\Psi(\mathbf{r}_1,\ldots,\mathbf{r}_N, t) \equiv \Psi(\mathbf{x}_N, t)$, its angular momentum $\mathbf{L}$ can be computed using the quantum mechanical formula [2]

$$\mathbf{L} = -i\hbar \int \Psi^*(\mathbf{x}_N, t) \sum_{j=1}^{N} [\mathbf{r}_j \nabla_j] \Psi(\mathbf{x}_N, t) d\mathbf{r}_1 \cdots d\mathbf{r}_N. \tag{4.1}$$

Upon integrating over all variables except $\mathbf{r}_j$ and making use of the definition of reduced density matrices, equation (4.1) can be transformed to the form (cf. equation (3.6) of [4] for the energy)

$$\mathbf{L} = -i\hbar \int \left\{ [\mathbf{r} \nabla] R_1(\mathbf{r}, \mathbf{r}') \right\}_{\mathbf{r}'=\mathbf{r}} d\mathbf{r}. \tag{4.2}$$

We assume that the system rotates around the $z$-axis and introduce spherical coordinates $r, \theta, \psi$ (we denote the second angle by $\psi$ to avoid confusion with the function $\varphi$). In this instance, only the component $L_z$ will be different from zero and equation (4.2) yields

$$L_z = -i\hbar \int \left[ \frac{\partial}{\partial \psi} R_1(\mathbf{r}, \mathbf{r}') \right]_{\mathbf{r}'=\mathbf{r}} d\mathbf{r}. \tag{4.3}$$

Hence it follows that $L_z \neq 0$, only if the function $\varphi(\mathbf{r})$ that determines $R_1$ depends upon $\psi$, and moreover $\varphi(\mathbf{r})$ is to be complex to give a real $L_z$. The density $\rho(\mathbf{r})$ of (2.9) will be independent of the angular coordinate $\psi$ and the symmetry will be axial, if $\varphi(\mathbf{r})$ is of the form

$$\varphi(\mathbf{r}) = \varphi(r,\theta)\, e^{il\psi}, \tag{4.4}$$

in which $l$ is an integer in order that $\varphi(\mathbf{r})$ be single-valued. Placing (2.2) with (4.4) in (4.3) and recalling the normalization of (2.10), we find the angular momentum of the system:

$$L_z = \hbar l N = \frac{\hbar}{m} l M. \tag{4.5}$$

First we take the simplest case $l = 1$. If we substitute (4.4) into the dimensionless equations of (2.16) and (2.18), drop the tilde and introduce $f(r,\theta) = \varphi(r,\theta)/\varphi_0$, we are led to

$$\frac{1}{r^2}\frac{\partial}{\partial r}\left(r^2 \frac{\partial f}{\partial r}\right) + \frac{1}{r^2 \sin\theta}\frac{\partial}{\partial \theta}\left(\sin\theta \frac{\partial f}{\partial \theta}\right) - \frac{f}{r^2 \sin^2\theta} + [\varepsilon - u(r,\theta)]\, f = 0, \tag{4.6}$$

$$\frac{1}{r^2}\frac{\partial}{\partial r}\left(r^2 \frac{\partial u}{\partial r}\right) + \frac{1}{r^2 \sin\theta}\frac{\partial}{\partial \theta}\left(\sin\theta \frac{\partial u}{\partial \theta}\right) = 4\pi f^2(r,\theta). \tag{4.7}$$

It can be shown that the solution for $f$ regular at $\theta = 0$ is of the form

$$f(r,\theta) = r\sin\theta\, g(r,\theta), \tag{4.8}$$

in which $g(r,\theta)$ is a regular function. From this it follows that the density $f^2(r,\theta)$ vanishes at $\theta = 0$, i.e., on the $z$-axis. Therefore the equilibrium rotating structure has the form of a ring.

As the function $f(r,\theta)$ is not spherically symmetrical, neither is the potential $u(r,\theta)$. For this reason, the function $f(r,\theta)$ does not break up into two factors, one of which depends on $r$ while the other on $\theta$, as happens with the solution of the Schrödinger equation in the case of a spherically symmetric potential. This considerably complicates numerical solution of equations (4.6), (4.7). The method employed for solving the equations is described in Appendix.

With this method, we succeeded only in obtaining a solution in which the function $g(r,\theta)$ of (4.8) has no zeros. However, other solutions, if any, should correspond to a higher energy as in section 3, in which case it is precisely the solution obtained that describes the equilibrium for $l = 1$. The calculated energy ($\tilde{E} = \tilde{\varepsilon}/3 = -9.514\cdot 10^{-3}$) is higher than for solution 1 of section 3 but it is less than for solution 2 (see table 1). Such an object, however, cannot evolve on its own into an immobile object relevant to solution 1, owing to the conservation of angular momentum. The plot of the density $f^2(r,\theta)$ as a function of $r$ in the equatorial plane is shown by curve 1 in figure 2. The behaviour of the density perpendicularly to the equatorial plane beginning at the point of the maximum is represented by curve 2 in the same figure. The spinning structure is a rather thick ring that has an inner radius of 3.3, an outer radius of 36.0 and a height of 42.6, if one defines the boundary as in section 3.

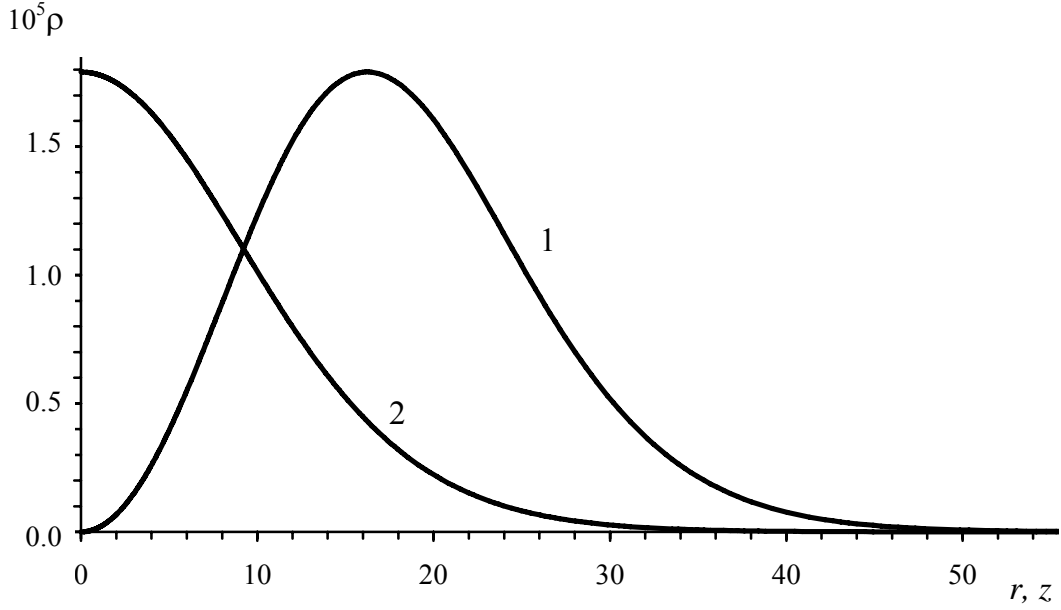

**Figure 2.** Density $\rho(r,\theta) = f^2(r,\theta)$ for a rotating structure. Curve 1: the $r$-dependence at $\theta = \pi/2$ ($z = 0$), curve 2: the $z$-dependence along a line parallel with the $z$-axis and passing through the point of the maximum density.

To elucidate the character of the rotation it makes sense to find the number density of current **j** (the number of particles crossing, on the average, unit area normal to the vector **j** per unit time). In the case of one particle the probability current density is given by [2]

$$\mathbf{i} = \frac{i\hbar}{2m}\left(\Psi\nabla\Psi^* - \Psi^*\nabla\Psi\right). \qquad (4.9)$$

Summing up over all particles one obtains, by analogy with (4.2), that

$$\mathbf{j} = \frac{i\hbar}{2m}\left[\frac{\partial}{\partial \mathbf{r}'}R_1(\mathbf{r},\mathbf{r}') - \frac{\partial}{\partial \mathbf{r}}R_1(\mathbf{r},\mathbf{r}')\right]_{\mathbf{r}'=\mathbf{r}}. \qquad (4.10)$$

When (2.2) and (4.4) are inserted into (4.10), one sees that only the component of the vector **j** tangent to the circumferences of revolution is nonzero, which is obvious in advance. On account of (4.8) the magnitude of the current density (in ordinary units) is found to be

$$j = \frac{\hbar}{m}lr\sin\theta\, g^2(r,\theta). \qquad (4.11)$$

It is more illuminating to introduce a mean velocity of particles **v** on a base of the hydrodynamic relation $\mathbf{j} = \rho\mathbf{v}$. In the present case $\rho = \varphi^2(r,\theta)$, and therefore (4.11) gives

$$|\mathbf{v}| \equiv v = \frac{l\hbar}{mr\sin\theta}. \qquad (4.12)$$

The divergence of $v$ as $\theta \to 0$ is fictitious because there are no particles at points where $\theta = 0$ according to (4.8), and $j = 0$ there. From (4.12) we see that the velocity decreases as the distance from the rotation axis $r\sin\theta$ increases. Therefore, the structure does not rotate as a solid since in the latter case the velocity should increase with the distance from the axis.

Here in fact we meet with an example of superfluidity. Rotating layers of the gas have different velocities without any dissipation of energy since the energy is stationary. This should not cause surprise, for the gas is in a state of Bose condensation. In [5] it was stressed that the phase with a condensate can be either superfluid or nonsuperfluid. In section 3 we had a condensate phase without superfluidity. The present condensate phase is superfluid.

We have considered in detail the case where $l = 1$ in (4.4). We discuss now in brief the case $l > 1$. In this case, one has $f(r,\theta) = r^l \sin^l\theta \, g(r,\theta)$ instead of (4.8). The radius of the ring grows with increasing $l$. The angular momentum increases as well according to (4.5).

## 5. Captured atmosphere

In order to consider the atmosphere of a celestial body composed of actual particles, it is necessary to take into account both gravitational and ordinary intermolecular forces. It is conceivable however that a solid devoid initially of atmosphere captures some amount of interstellar particles in its gravitational pull. If distances between the particles are sufficiently great (the relevant criterion is given in Introduction), the particles will interact only via the forces of gravity. Such a captured atmosphere will be considered in this section.

The gravitational field of the solid body plays the role of an external field $V^{(e)}(\mathbf{r})$ for the atmosphere's particles. In view of this, equation (2.4) with $s = 1$ yields now, instead of (2.12),

$$U(\mathbf{r}) = V^{(e)}(\mathbf{r}) + \int K(|\mathbf{r} - \mathbf{r}'|)\rho(\mathbf{r}') \, d\mathbf{r}'. \tag{5.1}$$

If the coordinate origin is placed at the centre of the body presumed to be spherical in shape, the relevant external potential is $V^{(e)}(\mathbf{r}) = -GmM_0 / |\mathbf{r}|$ where $M_0$ is the mass of the body. If we apply the Laplacian to both sides of (5.1), we arrive again at equation (2.13) since $\nabla^2 V^{(e)}(\mathbf{r}) = 0$ in the present case. Thus we have the same basic equations as before, namely, (2.7) and (2.13).

We introduce now dimensionless quantities. In the present problem there exists a natural parameter with a dimension of length, namely, the radius $R_0$ of the central body. Upon writing $\mathbf{r} = R_0 \tilde{\mathbf{r}}$ and $\varphi = \varphi_0 f$, we obtain equation (2.16) but now instead of (2.17) we have

$$\varepsilon = \frac{\hbar^2}{2mR_0^2}\tilde{\varepsilon}, \qquad U = \frac{\hbar^2}{2mR_0^2}u. \tag{5.2}$$

Equation (2.13) is brought to the form of (2.18) with

$$\varphi_0^2 = \frac{\hbar^2}{2Gm^3 R_0^4}. \tag{5.3}$$

Here $R_0$ is a given quantity, and thereupon $\varphi_0$ is defined unequivocally in contradistinction to (2.19) where the quantity $l_0$ was not fixed and we utilized freedom in choosing $l_0$ in order to reduce the normalization condition of (2.10) to the form of (2.20). Now we cannot reduce (2.10) to that form. If account is taken of the fact that now the integration in (2.10) is to be extended over the volume occupied by the atmosphere, which implies that $r \geq R_0$ (or $\tilde{r} \geq 1$), equation (2.10) assumes the form (hereinafter we omit the tilde over $r$)

$$I \equiv \int_{r \geq 1} \rho(\mathbf{r})\, d\mathbf{r} \equiv 4\pi \int_1^\infty f^2(r)\, r^2\, dr = \mu \frac{M}{M_0}. \tag{5.4}$$

Here we have introduced the mass of the atmosphere $M = mN$ and a dimensionless quantity

$$\mu = \frac{2Gm^2}{\hbar^2} M_0 R_0 = 2 \frac{\varepsilon_G}{\varepsilon_r}\left(\frac{R_0}{\lambda_C}\right)^2, \tag{5.5}$$

wherein $\varepsilon_G = GM_0 m/R_0$ is the magnitude of the gravitational energy of a particle at the surface of the body, $\varepsilon_r = mc^2$ is the relativistic energy of the particle, and $\lambda_C = \hbar/mc$ is its Compton wavelength. The second expression for $\mu$ may be helpful for comparing physical scales.

We shall consider a non-rotating atmosphere, which is equivalent to saying that we shall look for spherically symmetric and real solutions. In this case equations (2.16) and (2.18) take the form of (3.1). Equation (5.1) can be rewritten in the following dimensionless form

$$u(r) = -\frac{\mu}{r} - \frac{4\pi}{r}\int_1^r r'^2 f^2(r')\, dr' - 4\pi \int_r^\infty r' f^2(r')\, dr'. \tag{5.6}$$

One may verify directly that this expression satisfies the second equation of (3.1). Differentiating (5.6) and putting $r = 1$ yields

$$\left.\frac{du}{dr}\right|_{r=1} = \mu. \tag{5.7}$$

This provides a boundary condition for $du/dr$ independent of parameters of the atmosphere.

A general solution to differential equations of the type (3.1) contains four arbitrary constants. One of them is determined by virtue of (5.7). There remain three constants. Recall that in the situation considered in section 3 there were two constants characterising the solutions of interest. Therefore, in the present problem we need a supplementary condition that is to be found on physical grounds. For the sake of simplicity we shall confine ourselves to two limiting cases. In the first of them that can be called the complete non-adhesion of the atmosphere to the surface of the body, the atmosphere density vanishes at the body's surface, that is, $f(r) = 0$ at $r = 1$. In the second case that will be referred to as the complete adhesion of the atmosphere to the surface of the body, the atmosphere density is a maximum at the body's surface, that is, $df/dr = 0$ at $r = 1$.

In case $\mu$ is great and the atmosphere's mass is small ($M \ll M_0$), one can estimate the thickness of the atmosphere $H$ analytically. The integrals in (5.6) are of the same order of magnitude as the integral $I$ in (5.4). For this reason, if $M \ll M_0$, the terms with the integrals in (5.6) can be neglected in comparison with the first term on the right. Then, if we put $r = 1$ in the first equation of (3.1), we shall see that the assumed great value of $\mu$ can be compensated only by $\varepsilon$, that is to say, $\varepsilon \approx -\mu$ in this case. On the other hand, if $r \to \infty$, one can drop $u$ and the term containing $1/r$ in the same equation. The resulting equation can be solved readily to yield $f \propto \exp\left(-\sqrt{-\varepsilon}\, r\right) \approx \exp\left(-\sqrt{\mu}\, r\right)$. Hence it follows that the atmosphere's thickness may be estimated as $\tilde{H} \sim 1/\sqrt{\mu}$, which gives $H \sim R_0/\sqrt{\mu}$ in ordinary units.

Seeing that the solution depends essentially upon the parameter $\mu$, it is worthwhile to analyse its magnitude. We shall assume as before that the system, the atmosphere in the present context, is composed of hydrogen atoms. Then (5.5) yields

$$\mu = 0.336 \frac{M_0 R_0}{\text{g} \cdot \text{cm}}. \quad (5.8)$$

From this it is seen that for somewhat large central bodies the magnitude of $\mu$ proves to be very great. For example, in the case of a body akin to the Earth in mass and size, the value of $\mu$ is of the order $10^{36}$. With this $\mu$, the atmosphere's thickness would be $H \sim 6 \cdot 10^{-10}$ cm (at $T = 0$), which is two orders of magnitude less than atomic sizes. This last result is, of course, devoid of physical sense. Only values of $\mu$ of order unity or less, in which case $H$ will be sufficiently large, are of physical interest. From (5.8) we see that such values of $\mu$ are characteristic of bodies having a mass of about 1 g and a size of about 1 cm and under.

When solving the equations numerically, it is difficult to prescribe the atmosphere's mass $M$, i.e., to fix a value of the integral $I$ of (5.4) from the outset. For this reason, we prescribed a value of the derivative $f'(1)$ in the case of the complete non-adhesion of the atmosphere to the body's surface when $f(1) = 0$, or a value of $f(1)$ in the case of the adhesion when $f'(1) = 0$. The integral $I$ was evaluated afterwards. The procedure of solving the equations of (3.3) those of (3.1) are transformed to is analogous with the procedure described in section 3. The boundary condition of (5.7) was used, too. The parameters of the solutions obtained are presented in tables 2 and 3 for $\mu = 1$. Table 2 contains also values of the altitude $h_m$ above the body's surface where the function $f(r)$ reaches a maximum, and its value $f_m$ at the maximum. The thickness of the atmosphere $\tilde{H}$ is defined, by analogy with section 3, as the altitude where the density $f^{2}(r)$ decreases by a factor of 10 with respect to the density at the maximum. The energy of the atmosphere is calculated with the help of formulae, analogous to (2.22):

$$E = \frac{\hbar^4}{4Gm^4 R_0^3} \tilde{E}, \quad \tilde{E} = \tilde{\varepsilon} I - \tfrac{1}{2} \int\limits_{r\geq 1} u(\tilde{\mathbf{r}}) f^2(\mathbf{r}) \, d\mathbf{r}. \tag{5.9}$$

Table 2. Parameters of the captured atmosphere in the case of the complete non-adhesion of the atmosphere to the central body for $\mu = 1$.

| $f'(1)$ | $u(1)$ | $\tilde{\varepsilon}$ | $I$ | $h_m$ | $f_m$ | $\tilde{E}$ | $\tilde{H}$ |
|---|---|---|---|---|---|---|---|
| 0.01 | −1.005 | −0.1261 | $2.577 \cdot 10^{-2}$ | 1.75 | $5.03 \cdot 10^{-3}$ | $-6.528 \cdot 10^{-4}$ | 7.20 |
| 0.08158 | −1.235 | −0.2945 | 1.000 | 1.62 | $3.98 \cdot 10^{-2}$ | $-8.695 \cdot 10^{-2}$ | 5.94 |
| 0.1 | −1.319 | −0.3557 | 1.303 | 1.57 | $4.83 \cdot 10^{-2}$ | −0.1432 | 5.65 |
| 0.5 | −3.447 | −1.926 | 6.774 | 1.12 | 0.205 | −4.989 | 3.31 |
| 1 | −6.205 | −4.028 | 12.27 | 0.91 | 0.364 | −19.79 | 2.55 |
| 5 | −28.02 | −21.72 | 46.88 | 0.55 | 1.29 | −439.6 | 1.40 |

Table 3. Parameters of the captured atmosphere in the case of the complete adhesion of the atmosphere to the central body for $\mu = 1$.

| $f(1)$ | $u(1)$ | $\tilde{\varepsilon}$ | $I$ | $\tilde{E}$ | $\tilde{H}$ |
|---|---|---|---|---|---|
| 0.01 | −1.006 | −0.1922 | $1.726 \cdot 10^{-2}$ | $-5.280 \cdot 10^{-4}$ | 3.63 |
| 0.09659 | −1.383 | −0.4598 | 1.000 | −0.1255 | 2.97 |
| 0.1 | −1.405 | −0.4755 | 1.050 | −0.1379 | 2.95 |
| 0.5 | −5.193 | −3.298 | 7.263 | −9.270 | 1.56 |
| 1 | −11.73 | −8.543 | 16.01 | −57.16 | 1.10 |
| 5 | −101.0 | −87.72 | 119.6 | −4870. | 0.48 |

The plot of the function $f(r)$ for the $I = 1$ case is shown in figure 3. It should be mentioned that there are solutions in which $f(r)$ has zeros at $r \neq 1$.

Inspecting the tables one sees that the thickness of the atmosphere decreases as its mass increases (the integral $I$ increases). Therefore, if a solid body captures extra particles in its atmosphere, the atmosphere thickness diminishes. The values of $I$ presented in the tables show as well that the body is capable of capturing an atmosphere whose mass is many times the mass of the body. It is worth adding that, as the atmosphere's mass increases, the energy of the atmosphere decreases steeply, which amounts to saying that it is very favourable energetically for the body to augment the mass of its atmosphere.

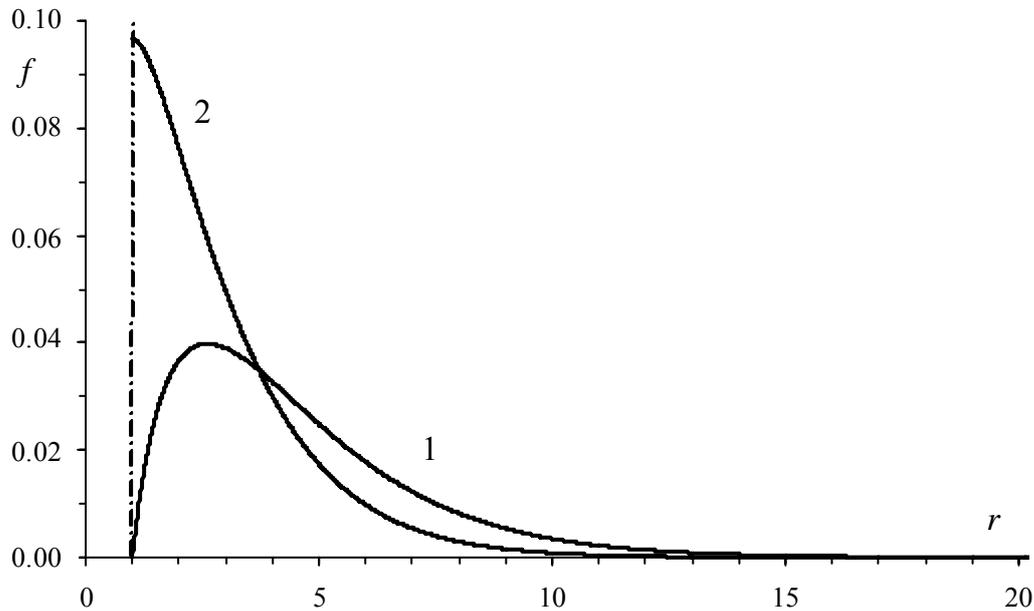

**Figure 3.** Function $f(r)$ for a captured atmosphere at $\mu = 1$ in the case where the mass of the atmosphere is equal to that of the body ($I = 1$). Curve 1: the complete non-adhesion of the atmosphere to the central body, curve 2: the complete adhesion.

It should be remarked that the tables were compiled with the sole purpose of demonstrating clearly the tendencies in mutual variations of different parameters, in case the forces of gravity alone act in the atmosphere. Some data may in reality correspond to densities at which short-range intermolecular forces should be taken into account.

## 6. Discussion and concluding remarks

The results of this paper show that quantum effects can play a peculiar role in many-body gravitating systems. Such a role of the quantum effects is due to the fact that the systems are implied to be at absolute zero of temperature in a state of Bose condensation. In this case there can exist spatially-bounded equilibrium structures both immobile and rotating. Depending on the number of particles, the dimensions of a structure may be enormous even on an astronomical scale or relatively small.

If one applies the results obtained in the paper to a system composed of actual atoms (or molecules), account must be taken of the fact that short-range interatomic forces are present as well, and the ground (equilibrium at $T = 0$) state of the system is determined principally by these forces alone. At the same time, distances between the particles can be so great that the interatomic forces will play no role, in case the system is in an excited state. Therefore, if the systems considered in this paper form (in the situations discussed in Introduction), they will in fact be in excited states. The lifetime of these states may however be very long, due to the following. In order that a system could pass into the ground state, a sufficient number of atoms of the system must approach one another simultaneously for distances that are less than $r_0$ when the short-range interatomic forces come into play. The probability of such an event is extremely small, if the mean distance between the particles is such as in the interstellar gas, let alone the intergalactic gas. A collision of only two atoms may lead solely to formation of a molecule, owing to exchange forces whose range is shorter than that of the van der Waals forces [2]. A large amount of interstellar atomic hydrogen indicates that even pair collisions of atoms are very rare in the Galaxy, not to mention triple collisions.

It should be emphasized that the starting equations of (2.7) and (2.8) are valid for any densities and any interaction, provided the interaction can be described by a two-body potential $K(|\mathbf{r}_1 - \mathbf{r}_2|)$. For this reason, one can elucidate how the leading role switches from the gravitational interaction to the intermolecular one as the density increases. One can look into the ground state as well. The problem, however, will be complicated substantially by the fact that, in the matter of high densities and short-range forces, the approximation of (2.11) fails and a pair correlation function is to be taken into account in $\rho_2(\mathbf{r}_1,\mathbf{r}_2)$. In fact, one must solve the whole hierarchy of equations given by (2.3)-(2.6). The situation is analogous with that occurring in the case of the classical BBGKY hierarchy [9] (see also [4]).

The results of the paper can be used to explain the first stages of formation of celestial bodies from interstellar and even intergalactic gases. Amongst a wealth of free atoms in a galaxy or in intergalactic space, a relatively small number of atoms can find themselves under conditions permitting the atoms to unite into a bound system. For example, in our Galaxy, interstellar hydrogen atoms may organize themselves into a bound system with a radius of about 10 km, for which no change in the average density is needed according to section 2. It is worthy of remark that it is not necessary that the bound system be made up of particles that are neighbours in space. In this connection the example of Cooper's pairs in a superconductor may be cited, where the size of a pair is several orders of magnitude greater that the average distance between conduction electrons.

The system formed will be in an excited state, and two lines of evolution are possible. Firstly, the system will pass into the ground state according to the ordinary quantum mechanical laws, which will be accompanied with decrease in size. This evolution may however be very slow in view of the long lifetime of the excited state discussed above. The second line of evolution follows from the results of this paper; namely, the system will capture other particles, which is not difficult and is beneficial energetically since $E \propto -N^3$ in view of (3.5). In this case the size of the system will decrease as well according to (2.24). In either case, as the density becomes sufficiently high, short-range intermolecular forces come into play. As a result, the very dilute gas gives rise to a compact body showing a tendency to enhance its mass. Following stages of evolution are beyond the scope of an equilibrium theory because the system will begin to heat up, and nonequilibrium processes will come into force.

The above scenario requires a sufficient number of initial particles. According to section 5, another scenario is possible in the matter of formation of celestial bodies. Minor bodies the size of small pebbles and even grains of sand can capture interstellar atoms one by one in their atmosphere. It is interesting to note that if a body of mass 1 g captures a hydrogen atom the gravitational Bohr radius $\hbar^2/(GM_0 m^2)$ will be 6 cm. As was stated in section 5 the mass of the atmosphere captured may far exceed the mass of the body itself. Again, as the atmosphere's mass increases, the dimensions of the atmosphere decrease, and the atmosphere becomes more and more dense. The subsequent processes will develop in a manner described above.

From section 4 it follows that the celestial structures formed can rotate. It is quite plausible that the structures are capable to capture only particles that meet certain conditions. If this is the case, the rotating structure will capture the particles that have a definite angular momentum and the angular momentum of the structure will grow in view of (4.5).

**Appendix**

Let us first transform equations (4.6) and (4.7) into a form helpful in numerical calculation. Observing that $\varepsilon < 0$ for bound states implied in the paper, we put $\varepsilon = -v^2$. Instead of (4.6) it is convenient to resort to equation (2.16) that can be recast as (hereafter we omit the tilde)

$$\nabla^2 f(\mathbf{r}) - v^2 f(\mathbf{r}) = u(\mathbf{r}) f(\mathbf{r}). \tag{A.1}$$

With use made of the relevant Green function equation (A.1) can be rewritten in the integral form:

$$f(\mathbf{r}) = -\frac{1}{4\pi} \int \frac{e^{-v|\mathbf{r}-\mathbf{r}'|}}{|\mathbf{r}-\mathbf{r}'|} u(\mathbf{r}') f(\mathbf{r}') d\mathbf{r}'. \tag{A.2}$$

We change the distance scale to $\bar{\mathbf{r}} = v\mathbf{r}$ and label functions that depend on $\bar{\mathbf{r}}$ with a bar drawn over the letter. Substituting (4.4) with $l = 1$ and taking into account the fact that the function $f(r,\theta) = \varphi(r,\theta)/\varphi_0$ must be real yields

$$\bar{f}(\bar{r}, \theta) = -\frac{1}{4\pi v^2} \int \frac{e^{-|\bar{\mathbf{r}}-\bar{\mathbf{r}}'|}}{|\bar{\mathbf{r}}-\bar{\mathbf{r}}'|} \cos(\psi - \psi') \, \bar{u}(\bar{r}', \theta') \, \bar{f}(\bar{r}', \theta') \, d\bar{\mathbf{r}}'. \tag{A.3}$$

If this equation is multiplied by $\bar{f}(\bar{r},\theta)$ and integrated over all space, we obtain

$$\nu^2 = -\frac{1}{4\pi}\int\frac{e^{-|\bar{\mathbf{r}}-\bar{\mathbf{r}}'|}}{|\bar{\mathbf{r}}-\bar{\mathbf{r}}'|}\cos(\psi-\psi')\,\bar{u}(\bar{r}',\theta')\,\bar{f}(\bar{r}',\theta')\bar{f}(\bar{r},\theta)\,d\bar{\mathbf{r}}'d\bar{\mathbf{r}}\bigg/\int\bar{f}^{\,2}(\bar{r},\theta)\,d\bar{\mathbf{r}}. \quad (A.4)$$

Instead of equation (4.7), directly from (2.12) we have

$$\bar{u}(\bar{r},\theta) = -\frac{1}{\nu^2}\int\frac{\bar{f}^{\,2}(\bar{r}',\theta')}{|\bar{\mathbf{r}}-\bar{\mathbf{r}}'|}\,d\bar{\mathbf{r}}'. \quad (A.5)$$

In the case of a rotating structure, $\bar{f}(\bar{r},\theta) = \bar{r}\sin\theta\,\bar{g}(\bar{r},\theta)$ by (4.8).

These last equations serve as basis for numerical solution by iteration. Upon starting from some initial $\bar{g}(\bar{r},\theta)$ and $\nu$, we find the initial function $\bar{u}(\bar{r},\theta)$ by (A.5). We introduce these quantities into the right-hand side of equations (A.3) and (A.4) and find new $\bar{g}(\bar{r},\theta)$ and $\nu$, and so on. The iteration continues until we reach a prescribed number of stable digits in the functions sought. A transformation analogous with (3.4) permits one to satisfy the normalization condition of (2.20).

It should be remarked that the integrals that enter into (A.3)-(A.5) are difficult for numerical evaluation because they are singular in view of $/\bar{\mathbf{r}}-\bar{\mathbf{r}}'/$ in the denominator. To sidestep the difficulty, we used Gegenbauer's addition theorem that enables one to express $\exp(-\alpha|\mathbf{r}-\mathbf{r}'|)/|\mathbf{r}-\mathbf{r}'|$ in terms of a series that contains modified Bessel functions and Legendre polynomials [10] (if $\alpha = 0$ this gives $1/|\mathbf{r}-\mathbf{r}'|$ of (A.5)). The angular dependence of all functions in (A.3)-(A.5) was represented in terms of expansions in the Legendre polynomials $P_k(\cos\theta)$. To handle the series numerically they were cut off with account taken of a desired accuracy. The appearing integrals over $\theta'$ lend themselves to analytical calculation. As a result, it remains only to evaluate numerically integrals over $\bar{r}'$.